\documentclass[twoside,twocolumn,english,aps,prl]{revtex4-1}

	\usepackage[usenames,dvipsnames]{xcolor}
	\usepackage{amsmath}
	\usepackage{amsfonts}
	\usepackage{amssymb}
    \usepackage{stmaryrd}
	\usepackage[colorlinks=true,citecolor=blue,linkcolor=red]{hyperref}
	\usepackage{graphicx}
	\usepackage{bbold}	
	\usepackage[makeroom]{cancel}		
	\usepackage{multirow}				
	\usepackage[normalem]{ulem}        
	\usepackage{array}
	\usepackage{comment}
	\usepackage{pdfpages}
	\makeatletter
	\AtBeginDocument{\let\LS@rot\@undefined}
	\makeatother

	\DeclareMathOperator{\tr}{tr}

\begin{document}

\title{Work Statistics and Adiabatic Assumption in Nonequilibrium Many-Body Theory}

\author{Yi Zuo$^{1,6}$}
\author{Qinghong Yang$^{2}$}\email{yqh19@mails.tsinghua.edu.cn}
\author{Bang-Gui Liu$^{1,6}$}
\author{Dong E. Liu$^{2,3,4,5}$}\email{dongeliu@mail.tsinghua.edu.cn}

\affiliation{$^{1}$Beijing National Laboratory for Condensed Matter Physics, and Institute of Physics, Chinese Academy of Sciences, Beijing 100190, China}
\affiliation{$^{2}$State Key Laboratory of Low Dimensional Quantum Physics, Department of Physics, Tsinghua University, Beijing, 100084, China}
\affiliation{$^{3}$Beijing Academy of Quantum Information Sciences, Beijing 100193, China}
\affiliation{$^{4}$Frontier Science Center for Quantum Information, Beijing 100184, China}
\affiliation{$^{5}$Hefei National Laboratory, Hefei 230088, China}
\affiliation{$^{6}$University of Chinese Academy of Science, Beijing 100046, China} 

\date{\today}


\begin{abstract}
Keldysh field theory, based on adiabatic assumptions, serves as an widely used framework for addressing nonequilibrium many-body systems.
Nonetheless, the  validity of such adiabatic assumptions when addressing interacting Gibbs states remains a topic of contention.
We use the knowledge of work statistics developed in nonequilibrium thermodynamics to study this problem. 
Consequently, we deduce a universal theorem delineating the characteristics of evolutions that transition an initial Gibbs state to another.
Based on this theorem, we analytically ascertain that adiabatic evolutions fail to transition a non-interacting Gibbs state to its interacting counterpart. However, this adiabatic approach remains a superior approximation relative to its non-adiabatic counterpart.
Numerics verifying our theory and predictions are also provided. 
Furthermore, our findings render insights into the preparation of Gibbs states within the domain of quantum computation.
\end{abstract}

\maketitle

\textbf{\em Introduction.}---The concept of adiabatic driving has been widely used in quantum physics, including Berry phase~\cite{berry}, zero-temperature many-body theory~\cite{fetter,coleman}, nonequilibrium many-body theory~\cite{keldysh,kamenev,stefanucci}, and adiabatic quantum computation~\cite{albash}. The adiabatic theorem~\cite{born,kato,griffiths,sakurai} guarantees the validity of the adiabatic assumption in studying those physics. Specifically, in the zero-temperature many-body (field) theory, the Gell-Mann-Low theorem~\cite{gellmann,fetter,coleman}, which is a specialization of the adiabatic theorem for interacting many-body systems, indicates that one can obtain the interacting ground state from a non-interacting ground state by adiabatically switching on the interaction Hamiltonian. Such a reduction greatly facilitate the treatment of interacting systems, as it makes the non-interacting Green's functions as the building blocks. 

Based on the Schwinger-Keldysh closed time formalism~\cite{schwinger,keldysh,konstantinov}, the nonequilibrium Green's functions serve as a useful framework for nonequilibrium many-body problems~\cite{kamenev,kamenev2,altland,sieberer,yqh,yqh2}. When considering nonequilibrium many-body systems, one is often staring from a Gibbs state at inverse temperature $\beta$, where the Hamiltonian is in the form of $H=H_0+\lambda_{1} H_1$ with $H_1$ being the interaction Hamiltonian and $\lambda_{1}$ the interaction strength (for interacting systems, $H_1$ does not commute with $H_0$). In order to deal with such an interacting initial state, Konstantinov and Perel'~\cite{konstantinov,stefanucci} proposed that one can regard the interacting Gibbs state as an evolution in the imaginary time axis and then treat it with the  imaginary time Matsubara formalism~\cite{matsubara}. 
Despite the mathematical rigor of this approach, it presents complexities due to the concurrent handling of both imaginary-time and real-time Green's functions. 
Moreover, the treatment of interacting Gibbs states using Matsubara formalism is already a difficult task. 
Therefore, a streamlined formalism, predominantly focusing on real-time Green's functions for nonequilibrium many-body problems, is advantageous. 
To this end, Keldysh suggested an approach wherein the interacting Gibbs state is considered as the final state of an evolution that initiates from a non-interacting Gibbs state at $t=-\infty$, with interactions being adiabatically switched on~\cite{keldysh,kamenev,stefanucci}.
Then, the building blocks reduce to non-interacting Green's functions and one only needs to concentrate on real times, as encountered in the zero-temperature many-body theory.

In contrast to the zero-temperature many-body theory, the validity of the adiabatic assumption within the nonequilibrium many-body framework remains an open question. 
Specifically, if an adiabatic evolution fails to transition a non-interacting Gibbs state to an interacting Gibbs state for a specified Hamiltonian~\cite{nikolai}, can the adiabatic assumption still be deemed a viable approximation when juxtaposed with alternative evolution protocols?
Furthermore, it is pertinent to explore the inherent characteristics of such evolution protocols capable of transitioning a non-interacting Gibbs state to its interacting counterpart. 
Insights from these properties might prove instrumental for Gibbs state preparation methodologies~\cite{ge,chowdhury,wangyl}. 
In subsequent discussions, we will refer to these evolution protocols as non-interaction-to-interaction (NI) evolution protocols.

In this work, we generally discuss these problems based on work statistics~\cite{jarzynski1,jarzynski2,crooks,jarzynski3,seifert,schuster,fei1,ortega} developed in the nonequilibrium thermodynamics community. Specifically, a universal theorem that holds for arbitrary quantum systems and determines the NI evolution protocol is derived. Generic calculations of work statistics show that the adiabatic evolution can not transition a non-interacting Gibbs state to an interacting Gibbs state. However, comparing with non-adiabatic evolution protocols, the final state of the adiabatic evolution is more close to the desired interacting Gibbs state up to an error of the order of $\mathcal{O}(\lambda_1^3)$. For non-adiabatic evolutions, the error is of the order of $\mathcal{O}(\lambda_1^2)$.

\textbf{\em Work statistics and Jarzynski equality.}---
Before examining the desired evolution protocol's properties, we briefly review work statistics and the Jarzynski equality~\cite{jarzynski1,jarzynski2,crooks,jarzynski3,seifert,schuster,fei1,ortega,zawadzki} within the context of quantum mechanical Hamiltonian systems.

\begin{figure}[t]
    \centering    \includegraphics[height=9.2cm]{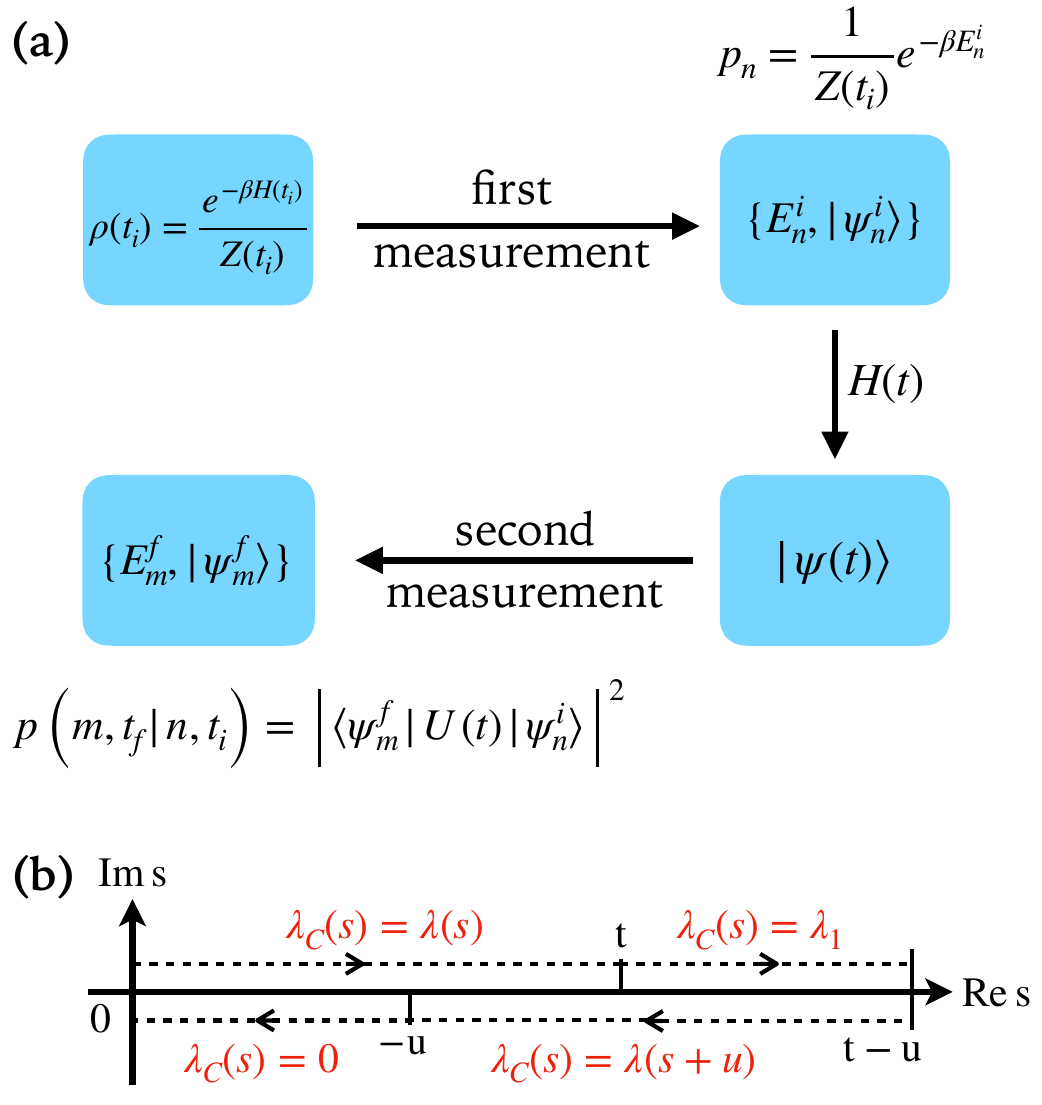}
    \caption{(a) Definition of work statistics through two energy measurements. (b) The contour $C$ used in calculating the characteristic function of work. The contour $C$ is divided into four parts for our case: $(i)$ $0\rightarrow t$: $\lambda_{C}\left(s\right)=\lambda\left(s\right)$; $(ii)$ $t\rightarrow t-u$ with $u<0$: $\lambda_{C}\left(s\right)=\lambda_{1}$; $(iii)$ $t-u\rightarrow t$: $\lambda_{C}\left(s\right)=\lambda\left(s+u\right)$; $(iv)$ $-u\rightarrow 0$: $\lambda_C(s)=0$. }
    \label{fig:dows}
\end{figure}

Ingredients of work statistics in quantum version can be defined through two energy measurements~\cite{talkner,fei2} as shown in Fig.~\ref{fig:dows} (a).  In the first measurement, the energy outcome is determined by the initial Gibbs state $\rho(t_i)=e^{-\beta H(t_i)}/Z(t_i)$ with $Z(t_i)=\tr[e^{-\beta H(t_i)}]$ being the partition function of the initial system. 
The first measurement can produce an eigenvalue $E_n^i$ of $H(t_i)$ with a probability $p_n = e^{-\beta E_n^i}/Z(t_i)$. 
Subsequent to this, the system transitions to the eigenstate $|\psi_n^i\rangle$ and evolves according to the time-dependent Hamiltonian $H(t)$ under the unitary evolution $U(t)$. At the final time $t_f$, another measurement results in the eigenvalue $E_m^f$ of $H(t_f)$ with a conditional probability $p(m,t_f|n,t_i) = \left|\langle\psi_m^f|U(t_f)|\psi_n^i\rangle\right|^2$. Here, $|\psi_m^f\rangle$ signifies the eigenstate of $H(t_f)$ corresponding to $E_m^f$. Consequently, the joint probability of obtaining measurements $E_m^f$ and $E_n^i$ is $p(m,t_f|n,t_i)p_n$.
The work is defined as the difference of two energy outcomes: $w=E_m^f-E_n^i$, and the probability of work $w$ should be 
\begin{equation}\label{eq:wd}
p(w)=\sum_{n,m}\delta\left[w-(E_m^f-E_n^i)\right]p(m,t_f|n,t_i)p_n.
\end{equation}

Having derived the work distribution, one can also define the characteristic function of work (CFW) through the Fourier transformation of $p(w)$:
\begin{equation}\label{eq:cfw}
\begin{split}
    \chi(u)&=\int dw e^{iuw}p(w)\\
    &=\frac{1}{Z(t_i)}\tr\left[U^{\dagger}(t_f)e^{iuH(t_f)}U(t_f)e^{-(iu+\beta)H(t_i)}\right].
\end{split}
\end{equation}
The CFW is a more convenient tool in studying nonequilibrium physics of quantum systems than the distribution $p(w)$. Remarkably, by setting $u=i\beta$ in the CFW, one can obtain the Jarzynski equality~\cite{jarzynski1}: 
\begin{equation}\label{eq:je}
\left\langle e^{-\beta w}\right\rangle=\frac{Z(t_f)}{Z(t_i)},   
\end{equation}
where $\langle\cdot\rangle$ without subscript is defined as $\langle\cdot\rangle=\int dw(\cdot)p(w)$, and $Z(t_f)=\tr[e^{-\beta H(t_f)}]$ is the partition function of a {\em hypothetical} system with Hamiltonian $H(t_f)$ in a Gibbs state at inverse temperature $\beta$. Note that the real system at $t_f$ is not necessarily at a Gibbs state. Hence, it is also desirable to ask that when the real system will be in a Gibbs state at $t_f$.

\textbf{\em Properties of the non-interaction-to-interaction (NI) evolution protocol.}---In the subsequent section, we will present our central theorem elucidating the properties of the NI evolution protocol. Prior to that, we introduce two lemmas essential for the theorem's proof. We provide only a succinct overview of the pivotal steps in the proof, with comprehensive details available in the Supplementary Information (SI)~\cite{supp}.

\textbf{Lemma 1.} The averaged work $\langle w\rangle\equiv\int dwwp\left(w\right)$ of an evolution from $t_i$ to $t_f$, can also be expressed as
$\langle w\rangle=\langle H\left(t_{f}\right)\rangle_{t_{f}}-\langle H\left(t_{i}\right)\rangle_{t_{i}}$, where $\langle H\left(t\right)\rangle_{t}\equiv\tr\left[\rho\left(t\right)H\left(t\right)\right]$.


\textbf{Lemma 2.} Suppose there exists an evolution protocol from $t_i$ to $t_f$, such that for all systems, the average work of the evolution satisfies $\langle w\rangle=\langle H(t_f)\rangle_G-\langle H(t_i)\rangle_{t_i}$, where 
\begin{equation*}
    \langle H(t_f)\rangle_G\equiv\frac{1}{\tr\left[e^{-\beta H(t_f)}\right]}\tr\left[e^{-\beta H(t_{f})}H(t_f)\right],
\end{equation*}
then the final state $\rho(t_f)$ is a Gibbs state with respect to $H(t_f)$ at inverse temperature $\beta$.


Having derived the above two lemmas, we are now in a position to prove our central theorem, which gives the property of the desired evolution protocol for {\em arbitrary} systems.

\textbf{Theorem 1.} Suppose there exists an evolution protocol from $t_i$ to $t_f$, such that for {\em all} systems, the evolution will drive the initial Gibbs state $\rho(t_i)=e^{-\beta H(t_{i})}/Z(t_i)$ to a final state (at $t_{f}$), which is a Gibbs state with respect to $H(t_f)$ at inverse termperature $\beta$. This evolution protocol exists {\em if and only if} the work distribution $p(w)$ is a delta function for all systems.

\textbf{Proof.} \textbf{a}). We first prove the sufficiency: If the work distribution is $p(w)=\delta(w-w_0)$ (For different systems, $w_0$ can be different), then the work of each trajectory should be the same, and equals to the average $\langle w\rangle$. According to the Jarzynski equality Eq.~\eqref{eq:je}, one has 
\begin{equation}
    \left\langle e^{-\beta w}\right\rangle=e^{-\beta w_0}=\frac{Z\left(t_{f}\right)}{Z\left(t_{i}\right)},
\end{equation}
where $Z(t_f)$ is the partition function of a {\em hypothetical} system with Hamiltonian $H(t_f)$ in a Gibbs state at inverse temperature $\beta$. Then
\begin{equation}
\begin{split}
    w_0&=-\frac{\partial}{\partial\beta}\ln\frac{Z\left(t_{f}\right)}{Z\left(t_{i}\right)}\\\Rightarrow\langle w\rangle&=-\frac{\partial}{\partial\beta}\ln\frac{Z\left(t_{f}\right)}{Z\left(t_{i}\right)}\\&=\langle H\left(t_{f}\right)\rangle_{G}-\langle H\left(t_{i}\right)\rangle_{t_{i}}.
\end{split}
\end{equation}
According to Lemma 2, we know that the state of the {\em real} system at $t_f$ is also the Gibbs state $\rho(t_f)=\exp{[-\beta H\left(t_{f}\right)]}/Z(t_f)$.\\ 
\textbf{b}). We then prove the necessity: If the state at $t_f$ is $\rho(t_f)=\exp{[-\beta H\left(t_{f}\right)]}/Z(t_f)$,
then according to Lemma 1, one has 
\begin{equation}\label{eq:aw}
\begin{split}
    \langle w\rangle&=\left\langle H\left(t_{f}\right)\right\rangle_{t_{f}}-\langle H\left(t_{i}\right)\rangle_{t_{i}}\\
    &=-\frac{\partial}{\partial\beta}\ln\frac{Z\left(t_{f}\right)}{Z\left(t_{i}\right)}.
\end{split}
\end{equation}
In addition, the Jarzynski equality will lead to
\begin{equation}\label{eq:voje}
    \begin{split}
        -\frac{\partial}{\partial\beta}\ln\left\langle e^{-\beta w}\right\rangle&=-\frac{\partial}{\partial\beta}\ln\frac{Z\left(t_{f}\right)}{Z\left(t_{i}\right)}.
    \end{split}
\end{equation}
Combinging Eq.~\eqref{eq:aw} and Eq.~\eqref{eq:voje}, one has 
\begin{equation}
    \begin{split}
        \langle w\rangle&=-\frac{\partial}{\partial\beta}\ln\left\langle e^{-\beta w}\right\rangle\\
        \Rightarrow\quad 0&=\left\langle w\left(\frac{1}{\langle e^{-\beta w}\rangle}e^{-\beta w}-1\right)\right\rangle.
    \end{split}
\end{equation}
Since this equation holds for {\em all} systems, the outer $\langle\cdot\rangle$ operation can be dropped, that is 
\begin{equation}
    \begin{split}
    0&=\frac{1}{\langle e^{-\beta w}\rangle}e^{-\beta w}-1\\
    \Rightarrow e^{-\beta w}&=\left\langle e^{-\beta w}\right\rangle\\
    \Rightarrow\quad\;\, w&=w_{0},
    \end{split}
\end{equation}
where $w_0$ is a constant but can be different for different systems. Therefore, the work distribution for the desired evolution protocol should be
\begin{equation}
    p(w)=\delta(w-w_0).
\end{equation}
$\hfill \boxempty$

The time-independent case is a trivial example of this theorem. In this case, the final Gibbs state is identical with the initial Gibbs state, and  the state after the first measurement only acquires a overall phase under the evolution provided by the time-independent Hamiltonian. Thus the trajectory work is simply $w=w_0=0$. 

Given the utility of the characteristic function of work (CFW) in facilitating analysis, we derive a corollary based on Theorem 1 in order to capture the property through the CFW.

\textbf{Corollary 1.} The logarithm of the characteristic function of work $\chi(u)$ for the evolution protocol given by Theorem 1 satisfies $\ln\chi(u)=iuw_0$, where $w_0$ is a real number.



\begin{figure*}[t]
    \centering    \includegraphics[height=7cm]{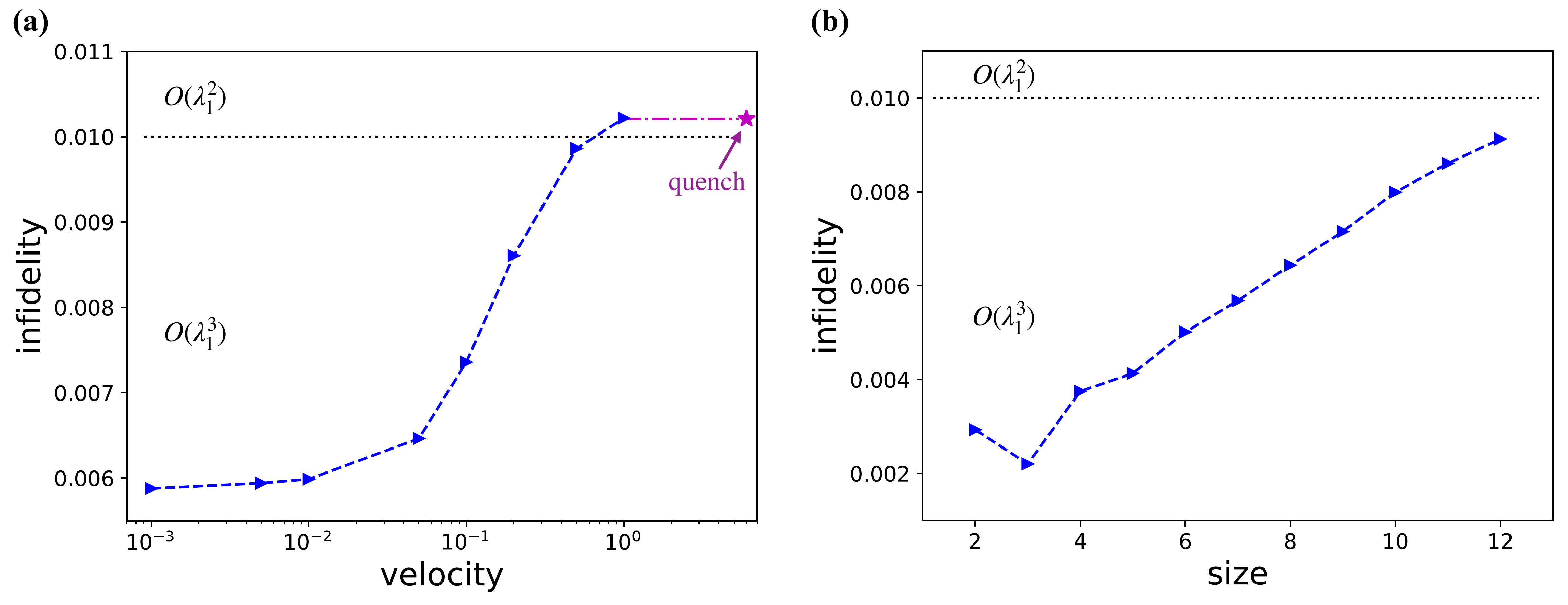}
    \caption{Infidelities ($1-$fidelity) between the interacting Gibbs state and final states of different evolution protocols. The parameters in common are chosen to be $J=2,\,\lambda_1 = 0.1,\,\beta = 1$. The total time for each evolution is $\lambda_1/0.001=100$.  (a) Infidelities for different evolution velocities of the linear driving $\lambda(t)=v t$. The size of the XXZ chain is chosen to be $N=11$. The final point marked as a star represents the quench evolution protocol, the velocity of which is regarded as $\infty$.  For velocities larger than $0.001$, the system will reach the full interaction before $t=100$, and then the system will continue to evolve with full interaction until time reaches $100$. (b) Infidelities for different sizes of the XXZ chain. The increasing velocity is chosen to be $v=0.2$ and the evolution time is also $t=100$. }  \label{fig:inf}
\end{figure*}

\textbf{\em Perturbative calculations of the characteristic function of work.}---For systems under an arbitrary nonequilibrium protocol, computing the CFW can be challenging. Nonetheless, when the full interaction strength, denoted as $\lambda_1$, within the Hamiltonian $H=H_0+\lambda_1H_1$ remains small, a universal formula for the CFW can be derived via perturbation theory~\cite{fei2}. Given the suitability of field theory techniques to weakly interacting quantum many-body systems, our focus predominantly lies within the perturbative domain.

Given the time evolution protocol where the interaction is gradually turned on from $0$ to $t$, we recognize that the exponential operators in Eq.~\eqref{eq:cfw} can all be treated as evolution operators along either real-time axis or imaginary-time axis. Thus, analogous to the Schwinger-Keldysh contour formalism, the CFW can be written in a contour-integral form~\cite{fei2}:
\begin{equation}
\begin{split}
\chi\left(u\right) = \left\langle\mathcal{T}_{C}\left[e^{-i\int_{C}ds\lambda_{C}\left(s\right)H_{1}^{I}(s)}\right]\right\rangle_0,
\end{split}
\end{equation}
where $\langle\cdot\rangle_0=\tr[(\cdot)e^{-\beta H_0}]/\tr[e^{-\beta H_0}]$, $\mathcal{T}_C$ is the contour-ordered operator with $C$ being the contour analogous to the Schwinger-Keldysh contour, and 
 $H_{1}^{I}\left(s\right)=e^{iH_{0}s}H_{1}e^{-iH_{0}s}$ is the interacting Hamiltonian in the interaction picture. The contour is divided into four parts as shown in Fig.~\ref{fig:dows} (b), according to the value of $\lambda_C(s)$. As the initial interaction strength is $0$, the contour only resides on the real-time axis.

The logarithm of $\chi\left(u\right)$,
known as the cumulant CFW, can be expanded through the cumulant correlation function~\cite{kubo}:
\begin{equation}
\begin{split}
\ln\chi\left(u\right)=\sum_{n=1}^{\infty}\int_{C}d\bar{s}_{1}\cdots d\bar{s}_{n}G_{c}\left(s_{1},\cdots,s_{n}\right),
\end{split}
\end{equation}
where $d\bar{s}_l\equiv ds_l\lambda_C(s)\theta_C(s_l-s_{l+1})$ with $\theta_C(s_l-s_{l+1})$ being the contour step function~\cite{fei2} and $\theta_C(s_n-s_{n+1})\equiv 1$, and $G_{c}\left(s_{1},\cdots,s_{n}\right)\equiv\left(-i\right)^{n}\langle H_{1}^{I}\left(s_{1}\right)\cdots H_{1}^{I}\left(s_{n}\right)\rangle_{c}$ is the $n$-point cumulant correlation function.

For non-adiabatic evolutions in our case, $\ln\chi(u)$ up to second order of $\lambda_1$ is given by:
\begin{equation}
\begin{split}
\ln\chi\left(u\right) = &iu\lambda_{1}\langle H_{1}\rangle_{c}+iu\lambda_{1}^{2}\int_{-\infty}^{\infty}\frac{d\omega}{2\pi}\frac{G^{>}_c(\omega)}{\omega}\\
&+\int_{-\infty}^{\infty}\frac{d\omega}{2\pi}\frac{1-e^{i\omega u}}{\omega^{2}}A(\omega)G^{>}_{c}(\omega)+\mathcal
{O}(\lambda_1^{3}
),
\end{split}\label{eq:Chi-2nd}
\end{equation}
where $G^>_c(\omega)$ is the Fourier transformation of $G^>_c(s_1-s_2)\equiv G_c(s_1,s_2)$, and $A(\omega)\equiv|\int_{0}^{t}ds\dot{\lambda}(s)e^{i\omega s}|^2$. Since $G_c^>(\omega)$ is a real function, the first and second term match Corollary 1, while the third term does not. For the adiabatic case, $t\rightarrow \infty$ and $\dot{\lambda}(s)\rightarrow 0$, then the third term containing $A(\omega)$ approaches to $0$. Thus, for adiabatic cases, $\ln\chi(u)$ is linear in $u$ when we keep terms up to $\mathcal{O}(\lambda_1^2)$, and then matches our theorem (or corollary). To see whether this holds for arbitrary order, we calculate $\ln\chi(u)$ up to $\mathcal{O}(\lambda_1^3)$ for the adiabatic case, and obtain
\begin{equation}
\begin{aligned}   \ln\chi\left(u\right)=&iu\lambda_{1}\langle H_{1}\rangle_{c}+i u\lambda_{1}^{2}\int_{-\infty}^{\infty}\frac{d\omega}{2\pi}\frac{G_{c}^{>}(\omega)}{\omega}\\
&+iu\lambda_{1}^{3}\int_{-\infty}^{\infty}\frac{d\omega_{1}}{2\pi}\frac{d\omega_{2}}{2\pi}\frac{G_{c}^{>}(\omega_{1},\omega_{2})}{i\omega_{1}(\omega_{1}+\omega_{2})}+\mathcal
{O}(\lambda_1^{4}
).
\end{aligned}
\end{equation}
where $G_{c}^{>}(\omega_{1},\omega_{2})$ is the Fourier transformation of $G^{>}_{c}(s_{1}-s_{3},s_{2}-s_{3})$, which is defined as 
\begin{equation}
\begin{split}
   &\quad G_c^>(s_1-s_3,s_2-s_3)\\
   &\equiv (-i)^3\langle H_1^I(s_1-s_3)H_1^I(s_2-s_3)H_1^I(0)\rangle_c\\
   &=G_c(s_1,s_2,s_3) 
   \end{split}
\end{equation}
Notice that $\mathcal{O}(\lambda_1^3)$ term does not match Corollary 1, as one can demonstrate that $G_c^>(\omega_1,\omega_2)$ is a complex function. Complete calculations can be found in SI~\cite{supp}.

Upon examining the universal criteria set by Theorem 1 (Corollary 1), we discern that neither adiabatic nor non-adiabatic evolution protocols can transition a non-interacting Gibbs state to its interacting counterpart. Notably, the logarithmic CFW for adiabatic protocols diverges from that of the NI protocol (shown in Corollary 1) to the order of $\mathcal{O}(\lambda_1^3)$, while for non-adiabatic protocols, the discrepancy occurs to the order of $\mathcal{O}(\lambda_1^2)$ (evident from the third term in Eq.(\ref{eq:Chi-2nd})). This suggests that although adiabatic evolution doesn't precisely achieve the desired state transition, it offers a superior approximation relative to non-adiabatic alternatives.


\textbf{\em Numerical verification.}---To verify our theory and predictions, we consider numerical results for a specific model---one-dimensional XXZ spin chain. The Hamiltonian reads
\begin{equation}    H_{xxz}=J\sum_{i=1}^{N-1}\left[\sigma^+_i\sigma^-_{i+1}+\sigma^-_{i+1}\sigma^+_i+\lambda(t) \sigma^z_i\sigma^z_{i+1}\right], 
\end{equation}
where $\lambda(t)$ controls the $zz$ interaction strength. For a given model, to check whether these two states are identical, one can directly compute the fidelity between the interacting Gibbs state and the final state after an evolution. Without loss of generality, we consider a linear driving protocol $\lambda(t) = v t$, where $v$ is the increasing velocity of the interaction strength. For a given system size, an evolution with larger $v$ is more non-adiabatic. We consider the small interaction regime, in which our perturbative calculation works. We finally arrive at Fig.~\ref{fig:inf} (a). One finds that larger velocities will lead to a final state more deviated from the target interacting Gibbs state. This confirms the result of our universal  analysis based on work statistics. Remarkably, in the nearly adiabatic regime (small $v$), the infidelity is approximately the order of $\mathcal{O}(\lambda_1^3)$, while in the non-adiabatic regime (large $v$), the infidelity is the order of $\mathcal{O}(\lambda_1^2)$. This result matches the prediction based on the logarithmic CFW. For smaller sizes with identical $J$ and $\lambda_1$, energy gaps will be larger, thus for a fixed increasing velocity, the evolution will be more adiabatic and one is expected to see smaller infidelities. This argument is confirmed by Fig.~\ref{fig:inf} (b). In addition, the infidelity in the adiabatic regime is still the order of $\mathcal{O}(\lambda_1^3)$.

\textbf{\em Conclusion.}---In summary, based on work statistics, we established a theorem elucidating the thermodynamic properties of an evolution transitioning an initial Gibbs state to another Gibbs state. In the realm of weak interactions, where the Keldysh field theory becomes particularly applicable, our theorem analytically demonstrates that adiabatic evolution does not precisely transition a non-interacting Gibbs state to its interacting counterpart. Nevertheless, when contrasted with non-adiabatic protocols, the resultant state from adiabatic evolution presents a closer approximation to the desired interacting Gibbs state.
Our numerical simulation confirms the theoretical predictions. 

\begin{acknowledgements}
{\em Acknowledgments.}---The work is supported by supported by the Innovation Program for Quantum Science and Technology (Grant No.~2021ZD0302400), and
the National Natural Science Foundation of China (Grant No. 11974198). \\
\end{acknowledgements}

Y. Zuo and Q. Yang contribute equally to this work.

\bibliography{WSAA_Ref}
\bibliographystyle{apsrev4-1}

\clearpage
\onecolumngrid

\includepdf[pages=1]{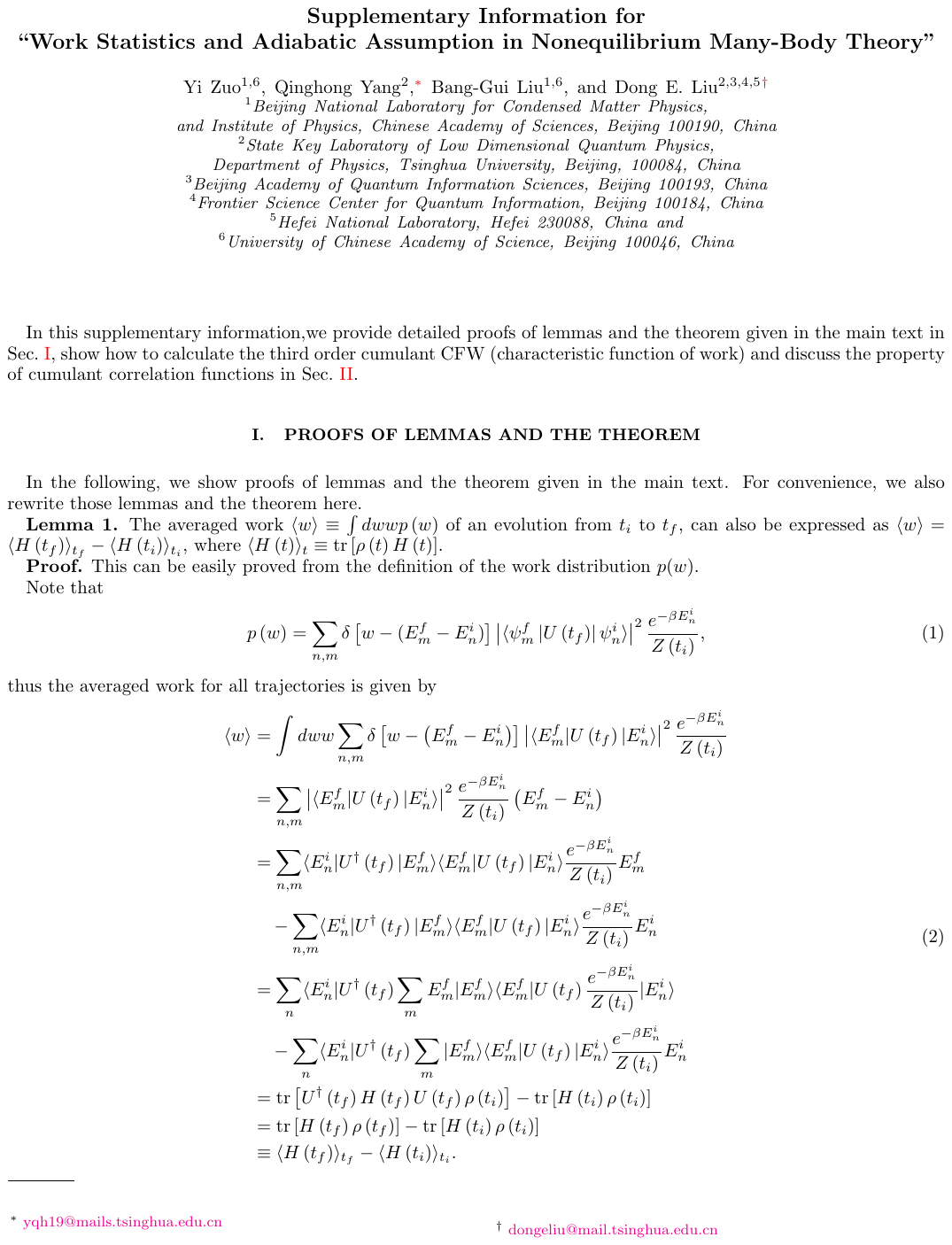}
\clearpage
\includepdf[pages=2]{WSAA_SI.pdf}
\includepdf[pages=3]{WSAA_SI.pdf}
\includepdf[pages=4]{WSAA_SI.pdf}
\includepdf[pages=5]{WSAA_SI.pdf}
\includepdf[pages=6]{WSAA_SI.pdf}
\includepdf[pages=7]{WSAA_SI.pdf}

\end{document}